\documentstyle[aps]{revtex}

\title{Chaotic Quantum Vortexes In A Weakly Non Ideal Bose Gas.
Thermodynamic Equilibrium And Turbulence }
\date{}
\begin{document}
\author{ { Sergey K. Nemirovskii}\footnote{E-mail: nemir@itp.nsc.ru}\\
\smallskip
{Institute for Thermophysics, Lavrentyeva,1, 630090 Novosibirsk, Russia}\\
 \ Makoto Tsubota\\
 \smallskip
 Department of Physics, Osaka City University, Osaka, Japan\\
 }
\maketitle
\begin{abstract}
    We study the stochastic behavior of a set of chaotic  vortex loops
appeared in imperfect Bose gas. Dynamics of Bose-gas is supposed
to obey Gross-Pitaevskii equation with additional noise satisfying
fluctuation-dissipation relation. The corresponding Fokker-Planck
equation for probability functional has  solution ${\cal
P}(\{{\psi }({\bf r})\})={\cal N}\exp (-H\left\{ {\psi }( {\bf
r)}\right\} /T),$ where $H\left\{ {\psi }({\bf r})\right\} $ is
the Ginzburg-Landau free energy. Considering vortex filaments as
topological defects of field ${\psi }({\bf r})$ we derive a
Langevin-type equation of motion of the line with the
correspondingly transformed stirring force. The respective
Fokker-Planck equation for probability functional ${\cal P}(\{
{\bf s}(\xi )\})$ in vortex loop configuration space is shown to
have a solution in the form of ${\cal P}(\{{\bf s}(\xi )\})={\cal
N}\exp (-H\left\{ {\bf s}\right\} /T),$ where ${\cal N}$ is the
normalizing factor and $H\left\{ {\bf s}\right\} $ is energy of
vortex line configurations. Analyzing this result we discuss
possible reasons for destruction of the thermodynamic equilibrium
and follow the mechanisms of transition to the turbulent state
\medskip
\newline \noindent PACS-number 67.40
\medskip
\end{abstract}

    Quantized vortices  (QV)  play a fundamental role in the properties
of superfluids and other  quantum systems.  Phase transitions,
kinetic properties and many other physical phenomena can be
effected by QV.  For that reason, they have been the object of
intense study for many years (for review and bibliography see
\cite {Don_book}). Studies of the QV dynamics have been most
successful in simple cases such as  vortex array in rotating
helium and vortex rings. These simple cases,however, are very
rare. Due to complex dynamics, initially straight lines or rings
can evolve into highly entangled  structures called vortex tangle
(VT), To describe any phenomena involving VT, some information
about its structure and statistics must be known. One of the most
popular ways to overcome that problem is to treat the vortex loops
as a kind of thermal excitation and to use the thermodynamic
methods. The first example of that way was the use of the  Landau
criterium for critical velocity where vortex energy and momentun
were used in the relation having purely thermodynamic sense. The
more recent example is the Kosterlitz-Thouless theory, as well as
its 3-D variant, currently being developed {for review and
bibliography see \cite{Williams}). Besides of vortices in the
condensed matter, the view of strings as a set of topological
defects being in thermodynamic equilibrium, is intensively
elaborating now in the field theory and in cosmology (see
e.g.\cite {Zurek}, \cite {Rivers}, \cite {Antunes}).

     On the other hand , numerous experiments involving counterflowing Hell, as
well as numerical simulation of vortex line dynamics
\cite{Don_book}, \cite{NF}, demonstrate that the system is not at
equilibrium. On the contrary, it possesses features of turbulent
phenomena. So, for example, the direct numerical  simulations
(\cite {Schwarz88}, \cite {npp93}, \cite {Brachet}, \cite
{Svistunov95}) demonstrate the presence of a cascade transfer of
energy both in space of the sizes of vortex loops, and in space of
Kelvin waves on every individual loop. Furthermore, in the
experiments involving vortices in quenched $^3$He and $^4$He (see
\cite{Krusius}, \cite{MacClintock}), the decay of vortex tangle
was observed rather than a thermalization as expected for normal
statistical systems. There are also serious discrepancies in the
statistical description of systems. As follows from theoretical
predictions founded on the study of dynamics quasi-equilibrium
topological defects \cite {Rivers}, decay of the vortex tangle
(rate of change of the density of vortices, or their total length
in unity of volume) occurs by the law $1/t $. Though this law
agrees with the decay of turbulent vortices (see \cite {Vinen}),
relaxation of quasi-equilibrium vortices, occurs by some orders
faster (\cite {Zurek}, \cite {Rivers}). Besides, to describe decay
of the turbulent tangle, Vinen has applied arguments used in the
theory of classical turbulence. Furthermore, distribution of
vortex loops with respect of their lengths $n (l) =dN (l) /dl $ in
turbulent helium obeys to the law $n (l) \propto l ^ {-4/3} $ (see
\cite {TAN2000}), while  the equilibrium vortices obey to the law
$n (l) \propto l ^ {-5/2} $ (see {\cite {Antunes2}}, \cite
{Vachaspati84}). The very important question is generation  of
turbulent vortices in the system. In the thermodynamic approach
vortexes arise due to thermal fluctuations of the order parameter,
or remain at prompt quench through the lambda line. The mechanisms
of appearing of turbulent vortices are not understood. Therefore,
the transition phenomena in the superfluid turbulence described by
the Vinen equation \cite {Vinen}, require introduction of
phenomenological "generative" term whose origin is not clear.

   Examples listed above  show, that chaotic vortex tangle, appearing in
quantum fluids can display both  the thermodynamic equilibrium
features and the extremely nonequilibrium, turbulent properties.
Naturally, there is a question: "What state, equilibrium or
turbulent, is realized in each concrete experiment, and what sorts
of the theory is required to use at description of one or another
phenomenon?" The answer to this question can give the
investigation devoted to study  of specific mechanisms,
responsible for establishment of thermodynamic equilibrium and
mechanisms featuring destroying the equilibrium and transition to
the turbulent state.
    In this work  we mainly concentrate on the first part of investigation
 formulated above and follow  how the vortices appeared in a weakly non-ideal Bose
gas are driven in thermodynamic equilibrium state. The second part
of  investigation is intended to be studied in further
publication. Here we only discuss, at a qualitative level, how
equilibrium state can be destroyed  and point out possible
mechanisms of transition to the turbulent state.

    As it was already mentioned, supposition about thermodynamic equilibrium  state can be
founded  by considering vortex loops as a subsystem submerged into
a thermostat with which they exchange energy. From the very
fundamental physical principles it follows that a set of vortex
loop should be driven in equilibrium state and, accordingly, to
obey the Gibbs statistics. The underlined physical field plays  a
role of a thermostat in this case. In case of a Bose condensate,
which we consider in given paper, the order parameter $ \psi ({\bf
x,} t)$ plays  the role of such field. Vortex filaments in this
study are described as crosses of the surfaces  where both real
and imaginary parts of the order parameter $\psi({\bf r,}t)$
vanish. Other excitations of the order parameter (phonons, and
rotons in case of real HeII) interact with an ensemble of vortex
loops, driving the latter  to thermodynamic equilibrium. In
accordance with the principle of maximal entropy, the probability
distribution function (functional) obeys the Gibbs formula.
      The given reasoning, however, bears a rather general character.
It gives no details concerning dynamics of the system since they
are hidden behind the temperature and chemical potential
definitions. Therefore it does not answer the question: " how the
system of vortices is driven in thermodynamic equilibrium? "
Without knowing  dynamical details we also can not to answer
another question: " how the thermodynamic equilibrium state for a
set of vortex loops is destroyed and the transfer to a strongly
nonequilibrium (turbulent) state occurs? " We stress that we` are
interested in analytical results, since there ia a number of
numerical works (see \cite {Antunes}, \cite {Antunes2}), where the
thermalization of vortices producing in equilibrium underlined
field was demonstrated.

      It is well known however that Gibbs distributuion can be
alternatively obtained in the frame of reduced models like kinetic
equations or the Fokker-Planck equation (FPE). These methods are
not as general as  a principle of maximal entropy, but instead
they allows us to follow the details of  how the Gibbs
distribution is established. In this study, FPE was used to
examine how chaotic vortex loops generated by the  thermodynamic
equilibrium Bose-Einstein condensates  reach equilibrium. This
study uses the Gross-Pitaevskii model \cite{Pit58} \cite{Gross61}
(for details and notations see \cite{Pismen99}).

    To achieve that goal we propose the following steps. First, we
state a problem of stochastic dynamics for the order parameter
$\psi ({\bf r,}t)$. Dynamics of quantity $\psi({\bf r,}t)$ is
supposed to satisfy the Gross-Pitaevskii equation \cite{Pit58}
\cite{Gross61}. To impose the stochastic behavior of a BEC, the
random stirring force $f({\bf x},t)$ was added to the right side
of the Gross-Pitaevskii equation. The random force obeys the
fluctuation-dissipation theorem. This guarantees that the Gibbs
distribution for probability functional ${\cal P }(\{{\psi }({\bf
r})\})={\cal N}\exp (-H\left\{ {\psi }({\bf r)}\right\} /T)$ is
followed. Here $H\left\{ {\psi }({\bf r)}\right\} $ is an energy
of particular configuration $\left\{ {\psi }({\bf r)}\right\} $ of
the order parameter, $T$ -temperature, and ${\cal N}$ is a
normalization factor.  Then, starting from that dynamics, we
derive an equation of motion for a vortex lines elements
cosidering the latter as zeroes of the order parameter. The random
force acting on the order parameter transforms into random
velocities for elements of the vortex line. Then we derive the
Fokker-Planck equation for probability distribution function
functional ${\cal P}(\{ {\bf s}(\xi )\})$ in a vortex loop
configuration space. Finally we show that the Fokker -Planck
equation is satisfied by the Gibbs distribution $\exp (-H\left\{
{\bf s}\right\} /T),$ where $H\left\{ {\bf s}\right\} $- is
functional of energy due to the vortex loops.  Analyzing this
result we discuss the possible reasons for destruction of
thermodynamic equilibrium and suggest the possible mechanisms of
transition to the turbulent state.
\section{Langevin equation for the order parameter}
    In this section we briefly describe the problem of the stochastic
motion of  Bose-condensate on the basis of Gross-Pitaevskii
model\cite{Pit58}\cite {Gross61}. In the frame of reference where
the normal velocity is absent the Gross-Pitaevskii equation reads
\begin{equation}
i\hbar \frac{\partial \psi }{\partial t}=(1-i\Lambda
)(-\frac{\hbar ^2}{2m} \nabla ^2\psi -\mu \psi +V_0\left| \psi
\right| ^2\psi ).  \label{NLSE}
\end{equation}
Here $\Lambda $ is (dimensionless) kinetic coefficient, $\mu $ is
chemical potential (energy per particle) and $V_0$ is the
potential describing an interaction of particles. The order
parameter $\psi $ is defined here so that
\begin{equation}
\psi =\sqrt{\rho _s}e^{i\phi }.  \label{psi_def}
\end{equation}
The quantity $\rho _s$ is the superfluid density, $\phi $ is
(dimensionless) phase, the flow velocity ${\bf v}$ is defined as
${\bf v}=(\hbar /m)\nabla \phi $. Rescaling the time $t\rightarrow
t(\hbar /m)$ and adding a random stirring force one can rewrite
\ref{NLSE} in the following form:
\begin{equation}
\frac{\partial \psi }{\partial t}=-\left( \Lambda +i\right)
\frac{\delta H(\psi )}{\delta \psi ^{*}}+f ({\bf x},t).
\label{Canonical1}
\end{equation}
Here $H\left\{ \psi \right\} $-is the Ginzburg-Landau free energy
functional, having the following form
\begin{equation}
H\left\{ \psi \right\} =\int d^3x\left[ \frac{\hbar
^2}{2m^2}\left| \nabla \psi \right| ^2-\frac \mu m\left| \psi
\right| ^2+\frac{V_0}{2m}\left| \psi \right| ^4\right].
\label{H(psi)}
\end{equation}
The thermal noise obeys the following fluctuation-dissipation
theorem
\begin{equation}
\left\langle f ({\bf x}_1,t_1)f ^{*}({\bf x}_2,t_2)\right\rangle
~=2k_BT\Lambda ~\delta ({\bf x}_1-{\bf x}_2)~\delta (t_1-t_2)~.
\label{fdt2}
\end{equation}
The stochastic problem described above is the well studied
problem, so called model A. It has a solution describing thermal
equilibrium, where probability of some configuration of $\psi
$-filed is proportional to $\exp (-H\left\{ { \psi }({\bf
r)}\right\} /T$. It immediately follows from the correspondent
Fokker-Planck equation\cite{Hohenberg}.
\section{Langevin equation for the vortex filament}
In this section we derive equation of the stochastic motion of
quantized vortex lines in Bose-condensate on the basis of the
problem formulated in \ref{Canonical1} and \ref{fdt2}. Vortices
are crosses zeroes of the both real and imagine parts of the
parameter, and the problem is just to describe the motion of
zeroes of the (complex) function $\psi$ varying in time and space.
That problem, as well as other more general problems of that kind
have been considered  many times by many methods, starting from
pioneering papers by Pitaevskii\cite{Pit61} and
Fetter\cite{Fetter66} (see also \cite{Pismen99}). We however
develop some new method, not too rigorous from mathematical point
of view, but convenient for the main purpose of the paper.

     Vortices in quantum fluids are very slender tubes (except
of the phase-transition region) and their dynamics is analogous to
that of the strings. This implies that we have to aim our efforts
at integrating out the radial degrees of freedom. It can be
reached by the following procedure.  As any motion of topological
defects, the vortex line dynamics is determined by the underlying
field theory. On the other hand, if one ignores for a while a
presence of other excitations except of the vortices one can say
that all motion of Bose-condensate is determined by a motion of
(all elements) of (all) quantum vortices. Hence, there is a mutual
correspondence and the order parameter $\psi ({\bf x},t)$ can be
considered as a functional of the whole vortex loop configuration
$\psi ({\bf x\mid s(}\xi ,t)).$ In particular, the temporal
dependance of field $\psi $ is connected with the motion of lines
and its rate of change (at some point ${\bf x}$ ) is expressed by
the following chain rule.
\begin{equation}
\frac{\partial \psi ({\bf x},t)}{\partial t}=\int\limits_\Gamma
\frac{\delta \psi ({\bf x},t)}{\delta {\bf s(}\xi ^{\prime
},t)}\frac{\partial {\bf s(} \xi ^{\prime },t)}{\partial t}d\xi
^{\prime }.  \label{dpsi/dt}
\end{equation}
Hereafter we will denote  integrals along the individual lines and
summation over all loops  by the index $\Gamma $. Let us rewrite
equation (\ref{Canonical1}) in the following form:
\begin{equation}
\frac{\Lambda -i}{\Lambda ^2+1}\frac{\partial \psi }{\partial
t}=-\frac{ \delta H(\psi )}{\delta \psi ^{*}}+~\frac{\Lambda
-i}{\Lambda ^2+1}f ( {\bf x},t). \label{Canonical2}
\end{equation}
Let us multiply equation (\ref{Canonical2}) by $\delta \psi
^{*}/\delta {\bf s(} \xi _0,t),$ where $\xi _0$ is some chosen
point on the curve. Adding the result to the complex conjugate and
integrating over whole space we have
\begin{eqnarray}
&&\int d^3{\bf x}\left( \frac{\Lambda -i}{\Lambda
^2+1}\frac{\partial \psi }{
\partial t}\frac{\delta \psi ^{*}}{\delta {\bf s(}\xi _0,t)}+\frac{\Lambda +i
}{\Lambda ^2+1}\frac{\partial \psi ^{*}}{\partial t}\frac{\delta
\psi }{ \delta {\bf s(}\xi _0,t)}\right) = \nonumber \\ &=&-\int
d^3{\bf x}\left( \frac{\delta H(\psi ,\psi ^{*})}{\delta \psi
^{*}} \frac{\delta \psi ^{*}}{\delta {\bf s(}\xi
_0,t)}+\frac{\delta H(\psi ,\psi ^{*})}{\delta \psi
^{*}}\frac{\delta \psi }{\delta {\bf s(}\xi _0,t)}\right) +~
\label{Can3} \\ &&+\int d^3{\bf x}\left( \frac{\Lambda -i}{\Lambda
^2+1}f ({\bf x},t) \frac{\delta \psi ^{*}}{\delta {\bf s(}\xi
_0,t)}+\frac{\Lambda +i}{\Lambda ^2+1}f ^{*}({\bf
x},t)\frac{\delta \psi }{\delta {\bf s(}\xi _0,t)} \right).
\nonumber
\end{eqnarray}
The  integral in the second line  of (\ref{Can3}) expresses a
chain rule for functional derivative $\delta H({\bf s})/\delta
{\bf s(}\xi _0,t)$, where $H( {\bf s})$ is the energy of
Bose-condensate expressed via vortex line position. Consequently
considering them to be very slender tubes (which is justified when
the radius of the curvature $R$ is much larger of the core size
$r_0$), and neglecting the part of energy associated with the
core, the quantity $H({\bf s})$ is just the kinetic energy of the
superfluid flow created by vortices (see e.g.
\cite{Bat},\cite{Saf})
\begin{equation}
H({\bf s})=\frac{{\rho }_s{\kappa }^2}{8\pi } \int\limits_\Gamma
\int\limits_{\Gamma ^{^{\prime }}}\frac{{\bf s}^{\prime }(\xi
){\bf s}^{\prime }(\xi^{\prime } )}{|{\bf s}(\xi )-{\bf s}(\xi
^{\prime })|}d\xi d\xi ^{\prime } .\label{H(s)}
\end{equation}
Here ${\bf s}^{\prime }(\xi )$ is a tangent vector,
{$\stackrel{\symbol{126}}{\kappa }$} is the quantum of circulation
equal to $2\pi \hbar /m$. Calculation of the functional derivative
$\delta H({\bf s})/\delta {\bf s(}\xi _0,t)$ is straightforward
and leads to result
\begin{equation}
\frac{\delta H({\bf s)}}{\delta {\bf s(}\xi _0,t)}={\rho
}_s{\kappa }{\bf s}^{\prime }(\xi _0)\times {\bf B(}\xi
_0).\label{dH/ds}
\end{equation}
Quantity  ${\bf B(}\xi _0)$ is the velocity of the vortex line
element expressed with  well known Biot-Savart law
\begin{equation}
{\bf B(}\xi _0)=-\frac{\kappa }{4\pi } \int\limits_{\Gamma
^{^{\prime }}}\frac{\left( {\bf s}(\xi _0)-{\bf s}(\xi ^{\prime
})\right) \times {\bf s}^{\prime }(\xi )}{|{\bf s}(\xi _0)-{\bf s}
(\xi ^{\prime })|^3}d\xi ^{\prime }. \label{Biot}
\end{equation}
Thus, we expressed the  integral in the second line  of
(\ref{Can3}) purely in terms of variable $ {\bf s(}\xi,t)$,
describing the vortex line configuration.

     Now we will treat the
integral in the lhs of equation (\ref{Can3}), which is the first
line of this relation . Substituting Rel. (\ref{dpsi/dt}) for
temporal derivative $\frac{\partial \psi }{\partial t}$ (and the
complex conjugate expression) into integrand we have
\begin{equation}
\frac{\Lambda -i}{\Lambda ^2+1}\int d^3{\bf x}\int\limits_{\Gamma
^{^{\prime }}}\left( \frac{\delta \psi ({\bf x\mid s(}\xi
,t))}{\delta {\bf s(}\xi ^{\prime },t)}\frac{\partial {\bf s(}\xi
^{\prime },t)}{\partial t}\right) \frac{\delta \psi ^{*}({\bf
x\mid s(}\xi ,t))}{\delta {\bf s(}\xi _0,t)}d\xi ^{\prime }+{\mbox
c.\mbox c.} \label{dpsi/dt2}
\end{equation}
Integral (\ref{dpsi/dt2}) depends on the whole vortex loop
configuration. Nevertheless this integral, (and other similar
integrals including squared gradients of the order parameter) can
be evaluated approximately in a general form, provided that the
line is slightly curved (see e.g.\cite{Pismen99}). The relevant
calculations, however, are rather cumbersome, and, to not block up
the basic text, we have taken out them in the Appendix. As follows
from the calculations given in the Appendix the term interesting
for us is a just sum of (\ref {s_dot_dif_1}) and (\ref
{s_dot_fin_app}), so the first line of equation ( \ref {Can3}) is
transduced in
\begin{equation}
\frac{2\pi \rho _s}{\Lambda ^2+1}\stackrel{\cdot }{\bf s}{\bf (}{
\xi } _0)\times {\bf s}^{\prime }(\xi _0)+\frac{2\pi \rho _s\sigma
\Lambda }{ \Lambda ^2+1}\stackrel{\cdot }{\bf s}{\bf (}{ \xi }_0).
\label{s_dot_fin}
\end{equation}
Thus, we again expressed the considered  term (first line of
(\ref{Can3})  in terms of variable $ {\bf s(}\xi,t)$, describing
the vortex line configuration.

     Let us now discuss the
rest term of equation (\ref{Can3}) including the random force $ f
({\bf x},t)$ (the third line). Consequently considering that all
motion of Bose-condensate is connected to motion of line ( We
remind, that we do not include acoustic effects), we have to
consider Langevin force $f ({\bf x},t)$ as some secondary quantity
stemming from random displacements of all elements of all loops.
This implies that the small casual variations of the order
parameter $ \delta\psi $ are connected to  the small casual
displacements of lines $ \delta {\bf s} $ with the help of the
following chain rule:
\begin{equation}
 \delta\psi ({\bf x},t)=\int\limits_\Gamma
\frac{\delta \psi ({\bf x},t)}{\delta {\bf s(}\xi ^{\prime
},t)}\delta {\bf s(}\xi ^{\prime },t)d\xi ^{\prime }.
\label{dz/dt}
\end{equation}
Furthermore, the casual small variations of $ \delta\psi $ and
small casual displacements of lines $ \delta {\bf s} $ (for some
interval of time $ \delta t $) are connected to random forces by
an obvious fashion
\begin{equation}
\delta {\bf s}(\xi ,t)={\bf f }(\xi ,t)\delta t,
\label{ds}
\end{equation}
\begin{equation}
\delta\psi ({\bf x},t)=f({\bf x},t)  \delta t. \label{dpsi}
\end{equation}
The reader should distinguish random force $f ({\bf x}, t) $
acting on the order parameter and random velocity $ {\bf f} (\xi,
t) $, acting  on the line on their arguments, besides $ {\bf f}
(\xi, t) $ - is, certainly, vector. Substituting (\ref {ds}),(
\ref {dpsi}) in (\ref {dz/dt}), and reducing on $ \delta t $, we
receive the following connection between random  force $f ({\bf
x}, t) $ and random velocity $ {\bf f} (\xi, t) $:
\begin{equation}
 f({\bf x},t)=\int\limits_\Gamma
\frac{\delta \psi ({\bf x},t)}{\delta {\bf s(}\xi ^{\prime
},t)}{\bf f}(\xi ^{\prime },t)d\xi ^{\prime }. \label{zfromz}
\end{equation}
Further we should substitute this expression (and also conjugate
expression) in the third line of equation (\ref {Can3}), that
gives
\begin{equation}
\frac{\Lambda -i}{\Lambda ^2+1}\int d^3{\bf x}\int\limits_{\Gamma
^{^{\prime }}}\left( \frac{\delta \psi ({\bf x\mid s(}\xi
,t))}{\delta {\bf s(}\xi ^{\prime },t)} {\bf f}(\xi ^{\prime },t)
\right) \frac{\delta \psi ^{*}({\bf x\mid s(}\xi ,t))}{\delta {\bf
s(}\xi _0,t)}d\xi ^{\prime }+{\mbox c.\mbox c.} \label
{dpsi/dtforce}
\end{equation}
This expression coincides with expression (\ref {dpsi/dt2}) with
an obvious replacement $ \stackrel {\cdot} {\bf s} ({\xi} _0)
\rightarrow {\bf f (} \xi _0, t) $. Applying technique used in the
Appendix we can directly write down the final result as the
similar to expression ( \ref {s_dot_fin})
\begin{equation}
\frac{2\pi }{\Lambda ^2+1}{\bf f (}\xi _0,t)\times {\bf s}^{\prime
}(\xi _0)+\frac{2\pi \Lambda }{\Lambda ^2+1}{\bf f (}\xi _0,t).
\label{f_fin}
\end{equation}
Note that our consideration does not relate intensities of the
random force $f ({\bf x}, t) $ acting on the order parameter and
random velocity $ {\bf f} (\xi, t) $, acting  on the line, which
will be done a bit later. Gathering together all transformed terms
of equation (\ref {Can3}), namely (\ref {dH/ds}), (\ref
{s_dot_fin}) and (\ref {f_fin}), we have (after cancellation by
$2\pi $)
\begin{eqnarray}
&&\frac 1{\Lambda ^2+1}\stackrel{\cdot }{\bf s}{\bf (}{ \xi
}_0)\times {\bf s}^{\prime }(\xi _0)+\frac{\Lambda \sigma
}{\Lambda ^2+1}\stackrel{ \cdot }{\bf s}{\bf (}{ \xi }_0)=\nonumber\\
&&={\bf s}^{\prime }(\xi _0)\times {\bf B(} \xi _0)+\frac
1{\Lambda ^2+1}{\bf f (}\xi _6,t)\times {\bf s}^{\prime }(\xi
_0)+\frac{\Lambda \sigma }{\Lambda ^2+1}{\bf f (}\xi _0,t)
\label{allterms}
\end{eqnarray}
Vector equation \ref{allterms} can be resolved up to velocity
$\stackrel{ \cdot }{{\bf s}_{\parallel }}{ \xi }_{0})$ along the
curve, which does not have physical meaning and can be removed by
suitable parameterization of label variable $\xi $.  Resolving
equation \ref{allterms} we arrive at
\begin{equation}
\stackrel{\cdot }{\bf s}{\bf (}{ \xi }_0)=\frac{1+\Lambda
^2}{1+\Lambda ^2\sigma ^2}{\bf B(}\xi _0)+\frac{(1+\Lambda
^2)\Lambda \sigma }{1+\Lambda ^2\sigma ^2}{\bf s}^{\prime }(\xi
_0)\times {\bf B(}\xi _0)+{\bf f (}\xi _0,t).  \label{master}
\end{equation}
Equation (\ref{master}) describes the motion of vortex line in
terms of line itself. It is the remarkable fact (not obvious in
advance) that noise ${\bf f (}\xi _{0},t)$ acting on the line is
also additive (does not depend on line variables).

     The last effort
which we have to do it is to ascertain both the statistic
properties of noise ${\bf f (}\xi _0,t)$ and its intensity.
Equation ( \ref{f_fin}) appeared as the result of transformation
of the third line of equation (\ref{Can3}) including random force
$f ({\bf x},t)$, so they are two equal vectors. Equating them we
have
\begin{eqnarray}
&&\frac \hbar m\frac{2\pi \rho _s}{\Lambda ^2+1}{\bf f (}\xi
_1,t_1)\times {\bf s}^{\prime }(\xi _1)+\frac \hbar m\frac{2\pi
\rho _s\Lambda }{\Lambda ^2+1}{\bf f (}\xi _1,t_1)=\label{f_equil_f}\\
&=&\int d^3{\bf x}_1\left( \frac{\Lambda -i}{ \Lambda ^2+1}f ({\bf
x}_1,t_1)\frac{\delta \psi ^{*}}{\delta {\bf s(}\xi
_1,t)}+\frac{\Lambda +i}{\Lambda ^2+1}f ^{*}({\bf
x}_1,t_1)\frac{\delta \psi }{\delta {\bf s(}\xi _1,t)}\right).
\nonumber
\end{eqnarray}
Taking a scalar productions of both parts of equation
(\ref{f_equil_f}) on themselves taking at another time $t_2$ and
point $\xi _2$ we get
\begin{eqnarray}
&&4\pi ^2\frac{\hbar ^2}{m^2}\frac{1+\Lambda ^2\sigma ^2}{\left(
\Lambda ^2+1\right) ^2}({\bf f }_{\eta _1}{\bf (}\xi _1,t_1){\bf f
}_{\eta _1}{\bf (}\xi _2,t_0)+{\bf f }_{\eta _2}{\bf (}\xi
_1,t_1){\bf f }
_{\eta _2}{\bf (}\xi _2,t_2))=  \label{intensity} \\
&=&\frac 2{\Lambda ^2+1}\int \int d^3{\bf x}_1d^3{\bf
x}_2\frac{\delta \psi ( {\bf x}_1,t_2)}{\delta {\bf s(}\xi
_2,t_1)}\frac{\delta \psi ^{*}({\bf x} _2,t_2)}{\delta {\bf s(}\xi
_2,t_2)}f ({\bf x}_1,t_1)f ^{*}({\bf x} _2,t_2). \nonumber
\end{eqnarray}
Here ${\bf f }_{\eta _4}$ and{\bf \ }${\bf f }_{\eta _2}${\bf \
}are components of random velocities in $\eta _{1},\eta _{2}$
directions in the normal (to the line) plain. We omit the rest
terms in the rhs foreseeing them to drop out after averaging.
Averaging relation (\ref {intensity}) and using the
fluctuation-dissipation theorem (\ref{fdt2}) we obtain
 \begin{equation}
\left\langle {\bf f }_{\eta _1}{\bf (}\xi _1,t_1){\bf f }_{\eta
_2} {\bf (}\xi _2,t_2)\right\rangle ~=\frac{k_BT}{\rho _s\pi
(\hbar /m)}\frac{ \left( \Lambda ^2+1\right) \Lambda \sigma
}{1+\Lambda ^2\sigma ^2}~\delta (\xi _1{\bf -}\xi _2)~\delta
(t_0-t_3)\delta _{\eta _{1}\eta _2}~. \label {fdt_line}
\end{equation}
    Thus, starting from dynamics of Bose-condensate ( equation
(\ref{Canonical1})) with fluctuation-dissipation theorem
(\ref{fdt2}) we derive equation ( \ref{master}) describing motion
of the vortex line in terms of line itself with the additive noise
obeying (\ref{fdt_line}). These relations complete a statement of
the  stochastic problem of quantized vortex dynamics under random
stirring velocity stemming  from the random force acting on the
underlying field. In the next subsection we demonstrate that this
problem has the equilibrium solution given by Gibbs distribution
$\exp (-H\left\{ {\bf s}\right\} /k_BT),$. Here $H\left\{ {\bf s
}\right\} $- is the functional of energy due to the vortex loop
(equation (\ref{H(s)})) and $ T$ is the temperature of
Bose-condensate.

\subsection{ The Fokker-Planck equation}
   To show it we, first, derive the Fokker-Planck equation
corresponding to Langevin type dynamics obeying  (\ref{master})
and (\ref{fdt_line}). Let us introduce the following probability
distribution functional
\begin{equation}
{\cal P}(\{{\bf s}(\xi )\},t)=\left\langle {\cal P}^M\right\rangle
=\left\langle \delta \left( {\bf s}(\xi )-{\bf s}(\xi ,t)\right)
\right\rangle .  \label{pdf}
\end{equation}
Here $\delta $ is a delta functional in space of the vortex loop
configurations. Averaging is fulfilled over an ensemble of random
forces. Rate of change of quantity ${\cal P}(\{{\bf s}(\xi )\},t)$
is
\begin{equation} \frac{\partial {\cal P}}{\partial
t}=\left\langle -\int d\xi \left\{ \frac{ \delta {\cal
P}^M}{\delta {\bf s}(\xi )}\stackrel{\cdot }{\bf s}_{\det
}\right\} \right\rangle -\int d\xi \left\{ \frac \delta {\delta
{\bf s}(\xi ) }\left\langle {\bf f }(\xi,t){\cal P}^M\right\rangle
\right\} . \label{FP1}
\end{equation}
Here $\stackrel{\cdot }{{\bf s}}_{\det }$ is the deterministic
part of velocity determined by ((\ref{master})). Omitting standard
details of derivation (see for details e.g. \cite{Zinn-Justin96})
we write the Fokker-Planck equation in a form
\begin{eqnarray}
&&\frac{\partial {\cal P}}{\partial t} +\int d\xi \frac \delta
{\delta {\bf s} (\xi )}\left\{ \left[ \frac{1+\Lambda
^2}{1+\Lambda ^2\sigma ^2}{\bf B(}\xi )+\frac{(2+\Lambda
^2)\Lambda \sigma }{1+\Lambda ^2\sigma ^2}{\bf s}^{\prime }(\xi
)\times {\bf B(}\xi )\right] {\cal P}+\right\}\nonumber \\
 &&+\int
d\xi\int d\xi ^{\prime }\left\langle {\bf f (}\xi {\bf )f (}\xi
^{\prime })\right\rangle \frac \delta { \delta {\bf s}(\xi
^{\prime })}{\cal P}\ =0.  \label{FP2}
\end{eqnarray}
Here $\left\langle {\bf f (}\xi {\bf )f (}\xi ^{\prime
})\right\rangle $ is the correlation function of random velocity
applied to the line and satisfying (\ref{fdt_line}). Our aim now
is to prove equation (\ref{FP2} ) to have a solution in the form
${\cal P}(\{{\bf s}(\xi )\})={\cal N}\exp (-H\left\{ {\bf
s}\right\} /T),$ where ${\cal N}$ is a normalizing factor. We
start the proof with the first term in integrand of (\ref{FP2}).
Using ( \ref{dH/ds}) and parameterization, that $\stackrel{\cdot
}{{\bf s}}{\xi }_0)$ is normal to the line, we write
\begin{equation}
{\bf B(}\xi )=\frac 1{{\rho }_s{\kappa }}{\bf s} ^{\prime }(\xi
)\times \frac{\delta H({\bf s)}}{\delta {\bf s(}\xi
_0,t)}.\label{B_trans}
\end{equation}
Let us rewrite the first term in integrand of (\ref {FP2}) in the
following form
\begin{equation}
\frac 1{{\rho }_s{\kappa }}\frac{1+\Lambda ^2}{1+\Lambda ^2\sigma
^2}\int d\xi \frac \delta {\delta {\bf s}(\xi )} \left( {\bf
s}^{\prime }(\xi )\times \frac{\delta H({\bf s)}}{\delta {\bf s(}
\xi ,t)}\exp (-H\left\{ {\bf s}\right\} /k_BT)\right).
\label{first_FP}
\end{equation}
Performing the functional differentiation and using a tensor
notation we rewrite (\ref {first_FP}) in the following form (we
omit the coefficient before the integral):
\begin{eqnarray}
&&\int d\xi \exp (-H\left\{ {\bf s}\right\} /k_BT)\epsilon
^{\alpha \beta \gamma }\frac{\delta {\bf s}_\beta ^{\prime }(\xi
)}{\delta {\bf s}_\alpha (\xi )}\frac{\delta H({\bf s)}}{\delta
{\bf s}_\gamma {\bf (}\xi ,t)}+ \label{conv}\\
 &&+\int d\xi \exp (-H\left\{ {\bf
s}\right\} /k_BT)\epsilon ^{\alpha \beta \gamma } \left( {\bf
s}_\beta ^{\prime }(\xi ) \frac \delta {\delta {\bf s}_\alpha (\xi
)}\frac{\delta H({\bf s)}}{\delta {\bf s}_\gamma {\bf (}\xi
,t)}\right)+\nonumber
\\
&&+(1/k_BT)\int d\xi \exp (-H\left\{ {\bf s}\right\}
/k_BT)\epsilon ^{\alpha \beta \gamma }{\bf s}_\beta ^{\prime }(\xi
)\frac{\delta H({\bf s)}}{\delta {\bf s} _\gamma {\bf (}\xi
,t)}\frac{\delta H({\bf s)}}{\delta {\bf s}_\alpha {\bf (} \xi
,t)}.\nonumber
\end{eqnarray}
The functional derivative $\delta {\bf s}_\beta ^{\prime }(\xi
)/\delta {\bf s}_\alpha (\xi )\propto \delta _{\beta \alpha }$,
therefore all of terms vanish due to symmetry. Thus, the
reversible term gives no contribution to the FP equation
(\ref{FP2}), one can say it is a divergent free term (see e.g.
\cite {Ma}). It is clear, that the said above refers only to a
case of thermal  equilibriums, i.e. it is valid only for the
Gibbs solution.

    The last term in the FP equation(\ref{FP2}) can be transformed by
use of fluctuation-dissipation relation for line (\ref{fdt_line})
in the following way:
\begin{eqnarray}
&&\int d\xi \int d\xi ^{\prime }\left\langle {\bf f (}\xi {\bf )f
(} \xi ^{\prime })\right\rangle \frac \delta {\delta {\bf
s}(\xi ^{\prime })} {\cal P}=\label{FP_diss}\\
 &&\int d\xi \int d\xi ^{\prime
}\frac{k_BT}\pi \frac{\left( \Lambda ^2+1\right) \Lambda \sigma
}{1+\Lambda ^2\sigma ^2}~\delta (\xi _1{\bf -}\xi _2)~\delta
(t_1-t_2)\delta _{\eta _1,\eta _2}~\frac \delta { \delta {\bf
s}(\xi ^{\prime })}\exp (-H\left\{ {\bf s}\right\} /k_BT)
\end{eqnarray}
Observing (\ref{dH/ds}) and (\ref {B_trans}) again  one convinced
himself that this term exactly compensate the second term in (\ref
{FP2}) as it should be.

    Thus, we have proved that the thermal equilibrium of
Bose-condensate results in the thermal equilibrium of a set of
vortex loops. In the next section we will review the basic steps
of our derivation. On that ground we will discuss how the thermal
equilibrium can be destroyed and consider (qualitatively) a
possible scenario of transition to a turbulent state.
\section{Analysis of solution. Possible scenario of transition to a turbulent state}
Analyzing our proof one can see that the following steps were
crucial:

     1. The additive white noise $f ({\bf x},t)$ acting on field
$\psi({\bf x},t) $ is transformed into the additive white noise
${\bf f }(\xi ,t)$ acting on vortex line position  ${\bf s}(\xi
,t)$.

    2. The intensity of noise $f(\xi ,t)$ expressed by (\ref{fdt_line})
exactly compensates the dissipative flux of probability
distribution functional in the Fokker-Planck equation (\ref{FP2}).
Furthermore, the compensation occurs in integrand, in other words
the detailed balance between pumping and dissipation takes place.

    3. The non-dissipative part of equation of motion of the line elements
gives no contribution to the flux of probability FP equation
(\ref{FP2}), it is divergent free (см. \cite {Ma}).

     Resuming the said above, it is possible to say, that by
the basic reason for an establishment of the thermal equilibrium
in set of the vortex loops is the detailed balance between pumping
and dissipation which occurs for all scales. It certainly agrees
with the basic principles of the statistical physics (see, for
example, \cite {LL6}). It should be realized that the rest part of
the equilibrium problem, namely evaluation of the partition
function (and,accordingly, all physics) basing on the Gibbs
solution ${\cal P}(\{{\bf s}(\xi )\})={\cal N}\exp (-H\left\{ {\bf
s}\right\} /T),$ is the far more difficult problem. The real
success is achieved mainly in numerical simulations \cite
{Antunes},\cite {Antunes2}. As for analytical studies, we would
like to point out a series of works by Williams with collaborators
who uses 3D variant of the Kosterlitz-Thouless theory (for review
and bibliography see \cite{Williams}). We appeal to results of the
cited works to continue our investigation. One of the most
striking results obtained in  these works is an observation that
in equilibrium there are vortex loops of appreciably  large sizes.
That result can be hardly foreseen by naive estimation that
probability of appearing of long loop is Boltzmann suppressed,
$\sim \exp(-{\sigma}l/kT)$, where $\sigma$ is energy per unit of
length. That implies, that, first, there is strong screening and,
second, vortex loops have a polymer like structure with folding of
line elements. Both phenomena reduce greatly the effective energy
per unit of line length. Actually distribution of loops with
respect to their lengths obey the law $n (l) \propto l ^ {-5/2} $
with some exponential factor (see \cite {Antunes2}).

     We think, that the presence of the long vortex loops is responsible for possible
destruction of thermodynamic equilibrium. Indeed, the small vortex
loops (rings), of order a wave length of phonons, have quite
definite  energy and momentum. Degrees of freedom due to an extent
of these objects are expressed weakly. In other words, the small
vortex  loops (ring) can be interpreted, as some kind of thermal
excitations, which just renormalize a density of the normal
component. Interaction with  other quasi-particles (phonons and
rotons) is reduced to usual collisions, that leads to a
renormalization of usual kinetic coefficients. Quantitatively this
renormalization is small, and can be appreciable only very close
to the lambda transition (see \cite {Williams}).

    As opposed to the small loops, the long loops, first, have a very large amount of
interior degrees of freedom. Secondly, the long weakly curved
filament interact with the normal component under the laws applied
to macroscopic rectilinear vortices (see, for example \cite
{Don_book}). Thus, the long and small loops interact with a
quantum fluid in different manners. The situation here is quite
similar to a 2D case, where an interaction with third sound arises
only from the vortex pairs of the large sizes, while small vortex
- antivortex pairs just renormalize the density of the superfluid
component (see \cite {AHNS80}). Because of a presence of long
vortices, the detailed balance can be easily destroyed. Indeed,
the strong enough large-scale perturbation (appeared e.g. due to
nonuniform external stream) can violate the equilibrium state. Two
scenarios, leading to the turbulent state are possible. First of
them is connected to development of nonlinear Kelvin waves on the
vortex line. Because of a nonlinear character of the equation of
motion, the large-scale perturbations on initially smooth line
interact, creating higher harmonics. They, in  turn, produce
higher harmonics, etc. In real space this process corresponds to
the tangling of a long vortex loop and formation of parts  with
very high  curvature (kinks or hairpins), which are subjected to
the strong dissipation. Thus, in space of the Kelvin  waves a
nonequilibrium state is established. This state is characterized
by the Kolmogorov flux  of curvature along scale in the
one-dimensional space describing the vortex lines (similar
scenarios are described in works \cite {Schwarz88}, \cite {npp93},
\cite {Svistunov95}).

    Other scenario, leading to violation of the equilibrium
state due to the presence of long loops, consists in the
following. In the presence of the normal component the long
vortexes are exposed to mutual friction, and, as a consequence to
the Magnus force. As a result, "ballooning" (in average) of the
vortex loops is possible. Enlargement of an amount of the long
loops displaces the law $n (l) \propto l ^ {-5/2} $, by slower
dependence. Uncompensated long vortex loops are exposed to
"redundant" reconnections, breaking down into  smaller loops. This
process can be interpreted, as the nonequilibrium Kolmogorov
cascade  in space of the loop sizes. The reader familiar with the
theory of superfluid turbulence, easily finds out, that the second
scenario coincides with the one, introduced by  Feynman for
nonequilibrium dynamics of the vortex tangle in the counterflowing
HeII (see reviews \cite {Don_book}, \cite {NF}, and also original
clause \cite {Feynman55}). The only difference is that we consider
the given process on a background of  thermal equilibrium, and the
cascade arises from deviation from  $n (l) \propto l ^ {-5/2} $
distribution.

    Certainly, both scenarios described above are not independent.
For example, at the moment of reconnection  there are strong
deformations on smooth lines, generating the Kelvin waves. On the
other hand, internal  dynamics of the vortex filaments influences
frequency and details of reconnection. Thus, analyzing the
received equilibrium solution, we have offered the scenarios,
according to which, the transition to strongly nonequilibrium,
turbulent state, is possible. We would like  to stress, that
realization of these scenarios is due to the presence  of very
long (actually macroscopic) vortex loops in the equilibrium
solution.

    Let's discuss now, what sorts of the macroscopic perturbations
are capable to destroy thermodynamic equilibrium. First of all, it
would be nonuniformities appearing during a flow of superfluid  in
channels. Other case is the counterflow of the normal and
superfluid components appearing in the heat flux. Superfluid
turbulence can  be also generated by intensive first and second
sounds \cite {NT82}. Furthermore, the vortices, which have arisen
in fast transition throw lambda-point at the first moments are in
nonequilibrium, and, probably, relax under the turbulent law
(compare \cite {Zurek}, \cite {Rivers}). There are other, more
exotic, mechanisms of destroying  of thermodynamic equilibrium,
for example, with use of the ion current.

     In Introduction we have mentioned about the problem of initial vorticity,
which was faced by Vinen in his description of the transient
superfluid turbulence (see \cite {Vinen}). In order to make his
equation to be self-consistent, he introduced the
phenomenological, "generative" term, whose origin is  not clear.
Under the scenarios described above, the transition in turbulent
state is just development of instability of long vortex loops with
respect to the large-scale perturbations. Since instabilities
develop very rapidly (usually exponentially),  a presence of a
plenty initial loops is not necessary. Thus, the phenomenological
"generative" term in the Vinen equation is just some composition
of a small amount of the long vortex loops and instabilities
increments. Of course, the confirmation of that point of view
requires the quantitative analysis  of the  problem of stability
of vortex loops, in the flowing (counterflowing) superfluids, this
is supposed to be studied in the further work.

\section{Conclusion}
Exploring the Langevin dynamics of a chaotic set of  quantized
vortex filaments appearing in a weakly nonideal  Bose gas, we have
followed the mechanism of establishment of thermodynamic
equilibrium. We have concluded, that the equilibrium state arises
due to the additive random forces, acting on the order parameter,
transforms into additive random velocities, acting on the lines in
a proper manner, to satisfy the fluctuation-dissipation relation
for the lines dynamics. It guarantees the detailed balance between
random velocity and dissipation. Noting, further, that in the
vortex tangle there present very long vortex loops, subjected to
large-scale perturbations, we offer the qualitative scenarios of a
destruction  of the thermodynamic equilibrium
and transition to the turbulent state. \\
The author is grateful to E. Sonin and G. Williams for fruitful
discussion. This work was partly supported by grants N 03-02-16179
from Russian Foundation for Basic research and grant N 2001-0618
from INTAS.
\renewcommand{\theequation}{A.\arabic{equation}}
\setcounter{equation}{0}
\appendix
\section*{Appendix}
A major contribution into integrals like (\ref{dpsi/dt2}) appears
from a vicinity of the vortex filament. More precisely these
integrals diverge, however a divergency has a weak logarithmic
character and usually the radius of curvature (or interline space)
is used as an upper cut-off. In practice this implies that we can
replace functional $\psi ({\bf x\mid s(}\xi ,t))$ by the static
vortex solution \cite{Pismen99} $\psi _v({\bf x}_{\perp })=\psi
_v({\bf s(}\xi _{cl},t)-{\bf x} )$, where $\xi _{cl}$ is the label
of point of the line closest to point $ {\bf x}$. It is clear that
vector ${\bf s(}\xi _{cl},t)-{\bf x=x}_{\perp }$ lies on the plain
normal to line and crossing point ${\bf s(}\xi _{cl},t)$. In
accordance with the said above, the main contribution into
integral comes from the points ${\bf x}$ remoted from the line at
the distances much smaller than $R$, namely $\left| {\bf s(}\xi
_{cl},t)-{\bf x}\right| \ll R$.
 The static solution $\psi _v({\bf
x}_{\perp })$  can be described as follows. The order parameter is
$ \psi _v({\bf x}_{\perp })=A(|{\bf x}_{\perp }|)\exp [i\,\varphi
]$, where $ A(|{\bf x}_{\perp }|)\propto |{\bf x}_{\perp }|$ for
$|{\bf x}_{\perp }|\ll $ $ r_0$ (core size) and $A(|{\bf x}_{\perp
}|) \approx \sqrt{\rho _s}$ for $|{\bf x} _{\perp }|\gg r_0$. The
phase $\varphi $ is just a polar angle. It is also useful to write
$\psi _v({\bf x}_{\perp })$ in Cartesian coordinates $\eta
_{1},\eta _{2}$ lying on the normal (wrt line) plain and crossing
points ${\bf s(}\xi _{cl},t)$. Then $\psi _v({\bf x}_{\perp
})=A(\sqrt{{\eta _{1}}^2+\eta _{2}^2})\exp [i\arctan \eta _{2
}/\eta _{1},]$. In vicinity of vortex  it is approximately $\psi
_v({\bf x}_{\perp })\propto \eta _{1}+i\eta _2.$ Thus, to evaluate
integral (\ref{dpsi/dt2}) we replace $\psi ({\bf x\mid s(} \xi
,t))$ with the function $\psi _v({\bf x}_{\perp })=\psi _v({\bf
s(}\xi _{cl},t)-{\bf x})$ and the integrating over $d^3{\bf x}$
with $d^2{\bf x}_{\perp}d\xi _{cl}$. Functional derivative $\delta
\psi ^{*}/\delta {\bf s(} \xi ^{\prime },t)$ should be evaluated
by the following rule:
$$ \frac{\delta \psi ({\bf x\mid s(}\xi ,t))}{\delta {\bf s(}\xi
^{\prime },t)} =\nabla _{\perp }\psi _v({\bf x}_{\perp })\;\delta
(\xi ^{\prime }-\xi _{cl}). $$
Likewise,
$$ \frac{\delta \psi ({\bf x\mid s(}\xi ,t))}{\delta {\bf s(}\xi
_0,t)}=\nabla _{\perp }\psi _v({\bf x}_{\perp })\;\delta (\xi
_0-\xi _{cl}). $$ Then the integration over $d\xi _{cl}$ and
$d\xi^{\prime} $removes delta-functions and the whole expression
is taken in point $\xi _0$. Finally we have
\begin{equation}
\frac{\Lambda -i}{\Lambda ^2+1}\int d^2{\bf x}_{\perp
}(\stackrel{\cdot } {\bf s}{\bf (}{\xi }_0)\nabla _{\perp }\psi
_v({\bf x}_{\perp }))\nabla _{\perp }\psi _v^{*}({\bf x}_{\perp
}).  \label{sdot1}
\end{equation}
Here $\stackrel{\cdot }{{\bf s}}({\xi }_0)$ is velocity of line at
point of observation. The second term in integrand of the first
line  of equation (\ref {Can3}) is just complex conjugate, which
is
\begin{equation}
\frac{\Lambda +i}{\Lambda ^2+1}\int d^2{\bf x}_{\perp
}(\stackrel{\cdot } {\bf s}{\bf (}{ \xi }_0)\nabla _{\perp }\psi
_v^{*}({\bf x}_{\perp }))\nabla _{\perp }\psi _v({\bf x}_{\perp
}).  \label{sdot2}
\end{equation}
The sum of (\ref{sdot1}) and (\ref{sdot2}) splits into parts with
sum of integrands and their difference. The difference term can be
rewritten with the help of famous formula of the vector algebra in
form
\begin{equation}
\frac i{\Lambda ^2+1}\int d^2{\bf x}_{\perp }\stackrel{\cdot }{\bf
s}{\bf (} { \xi }_0)\times (\nabla _{\perp }\psi _v^{*}({\bf
x}_{\perp })\times \nabla _{\perp }\psi _v({\bf x}_{\perp })).
\label{s_dot_dif}
\end{equation}
To evaluate (\ref{s_dot_dif}) let us rewrite the static solution
$\psi _v({\bf x} _{\perp })$ in local polar coordinates $(\left|
{\bf x}_{\perp }\right| ,\varphi )$ Vector production
 $\nabla _{\perp }\psi _v^{*}({\bf x}_{\perp })\times \nabla
_{\perp }\psi _v({\bf x}_{\perp })$ takes the form
\begin{equation}
 (\nabla _{\perp }\psi _v^{*}({\bf x}_{\perp
})\times \nabla _{\perp }\psi _v( {\bf x}_{\perp
}))=2i\frac{A(|{\bf x}_{\perp }|)A^{\prime }(|{\bf x}_{\perp
}|)}{\left| {\bf x}_{\perp }\right| }{\bf s}^{\prime }(\xi _0),
\end{equation}
And integral  enterring  (\ref{s_dot_dif}) is evaluated as
\begin{eqnarray*}
&&\int d^2{\bf x}_{\perp }(\nabla _{\perp }\psi _v^{*}({\bf
x}_{\perp })\times \nabla _{\perp }\psi _v({\bf x}_{\perp }))=\\
&&={\bf s}^{\prime }(\xi _0)\int 2i \frac{A(|{\bf x}_{\perp
}|)A^{\prime }(|{\bf x}_{\perp }|)}{\left| {\bf x} _{\perp
}\right| }2\pi \left| {\bf x}_{\perp }\right| d\left| {\bf
x}_{\perp }\right| =2\pi i\rho _s{\bf s}^{\prime }(\xi _0).
\end{eqnarray*}
Thus the difference part of (\ref{sdot1}) and (\ref{sdot2})
represented by (\ref {s_dot_dif}) takes the form
\begin{equation}
\frac{2\pi \rho _s}{\Lambda ^2+1}\stackrel{\cdot }{\bf s}{\bf (}{
\xi } _0)\times {\bf s}^{\prime }(\xi _0) \label{s_dot_dif_1}
\end{equation}
Let us now study contribution, which arises from the sum of
(\ref{sdot1}) and ( \ref{sdot2}). It is
\begin{equation}
\frac \Lambda {\Lambda ^2+1}\int d^2{\bf x}_{\perp }\left\{
(\stackrel{\cdot }{\bf s}{\bf (}{ \xi }_0)\nabla _{\perp }\psi
_v({\bf x}_{\perp }))\nabla _{\perp }\psi _v^{*}({\bf x}_{\perp
})+(\stackrel{\cdot }{\bf s} {\bf (}{ \xi }_0)\nabla _{\perp }\psi
_v^{*}({\bf x}_{\perp }))\nabla _{\perp }\psi _v({\bf x}_{\perp
})\right\}.  \label{s_dot_sum}
\end{equation}
The evaluation of (\ref{s_dot_sum}) is easily fulfilled in
Cartesian coordinates $\eta _{1}\,eta _{2}$ in ${\bf x}_{\perp }$
plain. Then the first term is
\begin{equation}
(\stackrel{\cdot }{\bf s}{\bf (}{ \xi }_0)\nabla _{\perp }\psi
_v({\bf x} _{\perp }))\nabla _{\perp }\psi _v^{*}({\bf x}_{\perp
})=\left( \stackrel{.} {\bf s}_1\frac{\partial \psi _v}{\partial
\eta _1}+\stackrel{.}{\bf s}_2 \frac{\partial \psi _v}{\partial
\eta _2}\right) \left( \frac{\partial \psi _v^{*}}{\partial \eta
_1}{\bf e}_1+\frac{\partial \psi _v^{*}}{\partial \eta _2}{\bf
e}_2\right). \label{sum1}
\end{equation}
Here ${\bf s}_{1},{\bf s}_{2}$ are components of the line velocity
in $\eta _{1}\eta _{2}$ directions, respectively, and ${\bf
e}_{1}, {\bf e}_{2}$ are unit vectors in these directions. Note
that expressions in first brackets of rhs of above relations are
scalars whereas the second ones are vectors. Adding ( \ref{sum1})
with complex conjugate, we obtain
\begin{eqnarray*}
&&{\bf e}_1\left( 2\stackrel{.}{\bf s}_1\frac{\partial \psi
_v}{\partial \eta _1}\frac{\partial \psi _v^{*}}{\partial \eta
_1}+\stackrel{.}{\bf s}_2\frac{
\partial \psi _v}{\partial \eta _2}\frac{\partial \psi _v^{*}}{\partial \eta
_1}
  \stackrel{.}{\bf s}_2\frac{\partial \psi _v}{\partial \eta
_1}\frac{
\partial \psi *_v}{\partial \eta _2}\right)+\\
 &+&{\bf e}_2\left( 2\stackrel{.}
{\bf s}_2\frac{\partial \psi _v}{\partial \eta _2}\frac{\partial
\psi _v^{*} }{\partial \eta _2}+\stackrel{.}{\bf
s}_1\frac{\partial \psi _v}{\partial \eta _2}\frac{\partial \psi
_v^{*}}{\partial \eta _1}+\stackrel{.}{\bf s}_1 \frac{\partial
\psi _v}{\partial \eta _2}\frac{\partial \psi _v^{*}}{
\partial \eta _1}\right).
\end{eqnarray*}
Combination
\begin{equation}
\frac{\partial \psi _v}{\partial \eta _2}\frac{\partial \psi
_v^{*}}{
\partial \eta _1}+\frac{\partial \psi _v}{\partial \eta _1}\frac{\partial
\psi _v^{*}}{\partial \eta _2}
\end{equation}
(the second and the third terms in both brackets) disappears for
static solution $\psi _{v}({\bf x}_{\perp })$, because the complex
conjugation is equivalent to change $\eta _{1}\rightarrow -\eta
_{2}$). Due to the symmetry the $\frac{\partial \psi _{v}}{
\partial \eta _{1}}\frac{\partial \psi _{v}^{\ast }}{\partial \eta _{1}}$ is
just $\frac{1}{2}\left| \nabla _{\perp }\psi _{v}({\bf x}_{\perp
})\right| ^{2}$ and integrating it gives the result:
\begin{equation}
\int d^2x_{\perp }\left| \nabla _{\perp }\psi _v({\bf x}_{\perp
})\right| ^2=2\pi \rho _s\ln \frac{a_0R}{r_0}=2\pi \rho _s\sigma
,\   \label{sigma}
\end{equation}
where $a_{0}$ is some number (weakly depending on $R$). Resuming
all said above we finally have  that the term sought transforms
into
\begin{equation}
\frac{2\pi \rho _s\sigma \Lambda }{ \Lambda ^2+1}\stackrel{\cdot
}{\bf s}{\bf (}{ \xi }_0). \label{s_dot_fin_app}
\end{equation}
The sum of expressions (\ref {s_dot_dif_1}) and (\ref
{s_dot_fin_app}) is the first line of the equation (\ref {Can3}),
written down in the terms of variable $ {\bf s (} \xi, t) $,
describing a vortex configuration.

\end{document}